# FAILURE OF CLASSICAL ELASTICITY IN AUXETIC FOAMS

J.H. Roh[1], C.B. Giller[2], P.H. Mott, and C.M. Roland

Naval Research Laboratory, Chemistry Division, Code 6120, Washington DC 20375-5342

*(August 24, 2012)*

Abstract

A recent derivation [P.H. Mott; C.M. Roland, *Phys. Rev. B* **80**, 132104 (2009).] of the bounds on Poisson's ratio, $\nu$, for linearly elastic materials showed that the conventional lower limit, −1, is wrong, and that $\nu$ cannot be less than 0.2 for classical elasticity to be valid. This is a significant result, since it is precisely for materials having small values of $\nu$ that direct measurements are not feasible, so that $\nu$ must be calculated from other elastic constants. Herein we measure directly Poisson's ratio for four materials, two for which the more restrictive bounds on $\nu$ apply, and two having values below this limit of 0.2. We find that while the measured $\nu$ for the former are equivalent to values calculated from the shear and tensile moduli, for two auxetic materials ($\nu < 0$), the equations of classical elasticity give inaccurate values of $\nu$. This is experimental corroboration that the correct lower limit on Poisson's ratio is 0.2 in order for classical elasticity to apply.

---

[1] National Research Council postdoctoral fellow.

[2] American Association for Engineering Education postdoctoral fellow.

I. Introduction

Poisson's ratio ($\nu$) is a constant that describes the transverse strain, $\varepsilon_{22}$ or $\varepsilon_{33}$, of an elastic body accompanying a longitudinal strain, $\varepsilon_{11}$

$$\nu = -\frac{\varepsilon_{22}}{\varepsilon_{11}} = -\frac{\varepsilon_{33}}{\varepsilon_{11}} \tag{1}$$

For a mechanically isotropic material, Poisson's ratio is unique, having but one value [1,2,3]. The classical theory of elasticity for infinitesimal linear strain links $\nu$ of an isotropic solid to the other elastic constants, including the moduli and Lamé constants [4,5]. Because of the appeal of representing strain as the sum of a volumetric and a deviatoric (shear) strain, the most common expression is in terms of the bulk, $B$, and shear, $G$, moduli [6]

$$G = B\frac{3(1-2\nu)}{2(1+\nu)} \tag{2}$$

Eq. (2), together with the requirement that these moduli are finite and positive, yields the well-known "classical" bounds on Poisson's ratio for isotropic materials

$$-1 < \nu < \tfrac{1}{2} \tag{3}$$

It is known that anisotropic materials, e.g., foams having a honeycomb or otherwise novel structure [7,8], can exhibit $\nu$ that deviate from these limits, but the behaviour of such materials deviates from linear elasticity, so that eq. (3) does not apply. However, as we have pointed out recently [4], almost without exception $\nu$ for isotropic materials does not fall below 0.2. The only materials for which $\nu < 0.2$ are those for which classical elasticity appears to be inapplicable.

The simplicity of expressing strains in terms of $B$ and $G$ does not elevate their significance above that of other elastic constants. Thus, expression can be derived for $\nu$ in terms of Young's modulus, the longitudinal modulus, the biaxial modulus, etc., and we have shown [4,5] that this leads to more restrictive limits on $\nu$ than eq. (3). Since all elastic constants are equally valid, the most restrictive bounds are the correct ones, since they do not yield discrepancies with any less restrictive limits. Accordingly, when classical elasticity applies, the limits on $\nu$ are found to be [4,5]

$$\tfrac{1}{5} < \nu < \tfrac{1}{2} \tag{4}$$

The left-hand-side of eq. (4) has the virtue of corresponding to data for real materials. This result was derived mathematically, using only the assumptions of classical elasticity. The implication is that when experiments yield values of Poisson's ratio smaller than 0.2, the equations of classical elasticity do not apply. This is not a trivial result, because the materials with $\nu < 0.2$ tend to be very hard (e.g., diamond [9], beryllium [10], and fused quartz [11]). This makes direct measurements difficult, whereby recourse is often made to using two elastic constants to calculate others. Given eq. (4), the fact that such calculations employ classical elasticity would make them invalid.

The purpose of this paper is to experimentally assess the applicability of classical elasticity whenever $\nu < 0.2$. We carry out elastic measurements on foams exhibiting isotropic auxeticity, originating in de-buckling of cell ribs which leads to large changes in transverse dimensions during longitudinal elongation [8,12]. We compare Poission's ratios measured directly on the foams, $\nu_{exp}$, with values calculated from the classical expression

$$\nu_{calc} = \frac{E}{2G} - 1 \tag{5}$$

where $E$ is Young's modulus. Although these foams have negative $\nu$, and thus deviate from eq. (4), they comply with eq. (3); thus, a comparison of $\nu_{exp}$ and $\nu_{calc}$ provides an experimental test of the proposition that relations such as eq. (5) are erroneous whenever $\nu < 0.2$.

We find herein that measurements of the dimensional changes during elongation yield values of $\nu$ that are significantly smaller than those calculated from eq. (5). The failure of this equation of classical elasticity confirms the prediction that classical elasticity is valid only when $\nu > 0.2$.

II. Experimental

The polyurethane (McMaster-Carr) was an open-cell foam with a density equal to 0.048g/cm$^3$. Auxetic foams were prepared by triaxial compression of rectangular samples (initially either 46.4 mm × 46.4 mm × 217.5 mm or 51.2 mm × 51.2 mm × 240 mm) in a mold (dimensions = 32 mm × 32 mm × 150 mm), followed by heating above the melting point of the hard segments (~105 °C). Upon removal from the mold following slow cooling to room temperature, the foams showed no significant expansion. To ensure no adhesion of the cell walls, the samples were stretched 20% in each of three orthogonal directions. We designate the as-received foam as PU1, and the auxetic samples as PU2 and PU3, where the number designates their respective volumetric strains of 1, 2, and 3. A solid elastomer sample was also prepared by curing 1,4-polyisoprene (Natsyn from Goodyear) with 2% by weight dicumyl peroxide for 30 minutes at 160 °C.

The tensile modulus and Poisson's ratio were measured on samples elongated on an Instron 5500R at a strain rate (= 0.002 s$^{-1}$) sufficiently slow to yield equilibrium values. Strains were determined from the displacement of fiducial marks on test specimens (initially 170 mm long and 36 mm wide), obtained from digitized photographs (Olympus E-PM1, 4032 × 3024 pixels). The pixel size was about 12 µm; the analysis software (Digplot; polymerphysics.net/software.html) provided about 1/10 pixel resolution. At least three sets of two marks each per longitudinal and transverse direction were used to calculate ν and the Young's modulus

$$E = \frac{\sigma_e}{\varepsilon_e} \quad (6)$$

where $\sigma_e$ and $\varepsilon_e$ are the respective engineering stress and strain.

The shear modulus was measured with a sandwich configuration, also using the Instron at a shear strain rate equal to 0.002 s$^{-1}$; test samples were 50 mm long × 4 mm wide × 4 mm thick. To verify these measurements, G was also determined using a torsion geometry on ring specimens (25.4 mm outer diameter and 11.7 mm inner diameter) with an ARES rheometer operating at the same low shear rate. The shear modulus is given by

$$G = \frac{ft}{lw\delta} \quad (7)$$

where $\delta$ and $f$ are the displacement, and force, respectively, and $l$, $w$, and $t$ are the respective sample length, width, and thickness. The shear strain $\gamma = \delta / t$.

III. Results

*Direct determination of elastic constants.*

PU1 is transversely isotropic, having a modulus 60% higher in the third dimension. Measurements were of the displacement of fiducial marks lying in the symmetric plane. The auxetic foams behave isotropically up to at least 5% strain. The deformation mechanism of the auxetic foams involves de-buckling of the cell ribs, which causes their modulus to be lower than that of the precursor material [8,12].

Figure 1 shows Poisson's ratio measured directly for the polyisoprene and the three foams. The uncertainty in the data arises mainly from our ability to resolve the fiducial images, although for the foams inhomogeneity of the inherent cell structure may also contribute. The polyisoprene has a homogeneous structure, and typically elastomers have Poisson's ratio within the range 0.49 and 0.5 [13]; that is, near the upper bound on $\nu$ in eqs. (3) and (4). We find no systematic variation for the polyisoprene over our range of strain measurements, obtaining $\nu_{exp} = 0.496$. The foams all show $\nu_{exp}$ that increases over the range of strains (*ca.* 1 – 5% elongation). For the PU1 measured in the isotropic plane, $\nu_{exp}$ increases about 10% with strain, and linear extrapolation to zero strain gives $\nu_{exp} = 0.20$. This is at the lower limit of the more restrictive range of eq. (4). Over this same range of strains the auxetic foams show an increase in $\nu_{exp}$ of about 20%. We extrapolate to zero strain by a linear fit to the data, obtaining $\nu_{exp} = -0.70$ and $-0.65$ for PU2 and PU3, respectively. The more compressed foam has a smaller (absolute value) of Poisson's ratio. Similar results were reported previously for polyurethane foams [12]. For both auxetic materials herein Poisson's ratio is within the conventional limits of classical elasticity (eq. (3)), but beyond the more restrictive range (eq. (4)) posited in refs. [4,5]. All values of Poisson ratios determined by direct measurement of longitudinal and transverse deformations, $\nu_{exp}$, are tabulated in Table 1, along with the uncertainties.

Figure 2 displays Young's moduli for the foams; it was constant up to a few percent tensile strain for all materials. Also shown are the shear moduli, which showed some dependence on strain. The shear moduli measured by torsional rheometry were consistent with these data. Regression yields the zero strain values given in Table 1. Note that PU3, which has the greater volume compression, has a larger shear modulus than PU2. Since the mechanical response involves de-buckling of the foam, there is no certainty that the measured behaviour extrapolates smoothly to zero strain. Thus, the limit of error on *G* for the foams (Table 1) is taken as the difference between the value determined by extrapolation to zero strain and the value measured for the lowest strain.

*Poisson's ratio calculated from elasticity equations*

Poisson's ratio was calculated from the equation of classical elasticity (eq. (5)), using the values determined for the shear and Young's moduli. For polyisoprene and PU0, whose $\nu_{exp}$ fall at the upper and lower bounds of the more restrictive range (eq. (4)), the difference between the calculated and experimental values is less than 1%; that is, the agreement is well within the experimental uncertainties. This is expected, and does not prove the superiority of eq. (4) over eq. (3), but rather affirms the validity of our experimental methods.

The situation is different for the two auxetic foams. For both $\nu_{exp} < 0.2$, so that equations such as eq.(5) are inaccurate if eq. (3) represents the range of validity of classical elasticity. We find for both PU2 and PU3, $\nu_{calc}$ underestimates the measured Poisson's ratio, by an amount (about 10%) that exceeds the experimental uncertainties (which are less than 3%).

IV. Summary

Classical elasticity applies to small deformations for which the mechanical response is linear (e.g., strain energy quadratic in the strain), and the behaviour is elastic [2,12]. Direct determinations of Poisson's ratio are rare, since the errors are large; moreover, small absolute values of $\nu$ exacerbate the difficulties. Thus, $\nu$ is usually calculated from two other elastic constants. Unfortunately, for small $\nu$, as found for very hard materials, is the situation for which the classical equations cannot be used. The correct range of applicability of classical elasticity derived in recent analyses [4,5] is that given by eq.(4).

In this work we measured Poisson's ratio directly for four materials. The behavior of a homogeneous elastomer and a foam both comply with even the stricter limits of eq. (4); for these two materials $\nu_{calc}$ and $\nu_{exp}$ are in agreement. For the two auxetic foams, however, Poisson's ratio is less than the lower bound of eq. (4); thus, $\nu_{calc}$ will equal $\nu_{exp}$ only if the conventional range (eq.(3)) is the correct one. However, we find for both materials that classical elasticity calculations significantly underestimate the experimentally measured Poisson's ratio. These results corroborate the mathematical derivation [4,5] underlying eq. 4. The use of the equations of classical elasticity to calculate $\nu$ or other elastic constants is in error for any material for which Poisson's ratio is smaller than 0.2.

V. Acknowledgements

We thank Daniel Fragiadakis for helping to fabricate the molds and useful discussions. J.H. Roh and C.B. Giller acknowledge respective postdoctoral fellowships from the National Research Council and the American Society for Engineering Education. The work was supported by the Office of Naval Research.

VI. References

[1] G.N. Greaves, A.L. Greer, R.S. Lakes, and T. Rouxel, *Nature Matl.* **10**, 823 (2011).

[2] A.E.H. Love, *A Treatise on the Mathematical Theory of Elasticity*, Dover, New York; 1966.

[3] N.W. Tschoegl, W.G. Knauss, and I. Emri, *Mech. Time-Dep. Matl.* **6**, 3 (2002).

[4] P.H. Mott; C.M. Roland, *Phys. Rev. B* **80**, 132104 (2009).

[5] P.H. Mott and C.M. Roland, **arXiv:1204.3859** (2012).

[6] R.S. Lakes and R. Witt, *Int. J. Mech. Eng. Edu.* **30**, 50 (2000).

[7] K.E. Evans and A. Alderson, *Adv. Matl.* **12**, 617 (2000).

[8] R. Lakes, *Science* **235**, 1038 (1987).

[9] C.A. Klein and G.F. Cardinale, *Diamond Relat. Matl.* **2**, 918 (1993).

[10] A. Migliori, H. Ledbetter, D.J. Thoma, and T.W. Darling, *J. Appl. Phys.* **95**, 2436 (2004).

[11] T. Rouxel, *J. Am. Ceram. Soc.* **90**, 3019 (2007).

[12] J.B. Choi and R. S. Lakes, *J. Mat. Sci.* **27**, 4678 (1992).

[13] M.L. Anderson, P.H. Mott, and C.M. Roland, *Rubber Chem. Technol.* **77**, 293 (2004).

Table 1. Elastic constants and Poisson's ratio.

| | polyisoprene | PU1 | PU2 | PU3 |
|---|---|---|---|---|
| $E(\varepsilon=0)$ | 1.145 ± 0.002 | 0.1450 ± 0.0003 | 0.0451 ± 0.0002 | 0.0502 ± 0.0002 |
| $G(\gamma=0)$ | 0.382 ± 0.003 | 0.0601 ± 0.0005 | 0.0550 ± 0.0002 | 0.0640 ± 0.0007 |
| $G(\gamma=0.005)$ | --- | --- | 0.0524 ± 0.0011 | 0.0605 ± 0.0010 |
| $\nu_{calc}\ \gamma=0$ | 0.499 ± 0.014 | 0.206 ± 0.012 | −0.590 ± 0.020 | −0.608 ± 0.023 |
| $\nu_{exp}$ | 0.496 ± 0.006 | 0.204 ± 0.006 | −0.699 ± 0.008 | −0.650 ± 0.015 |

FIGURES

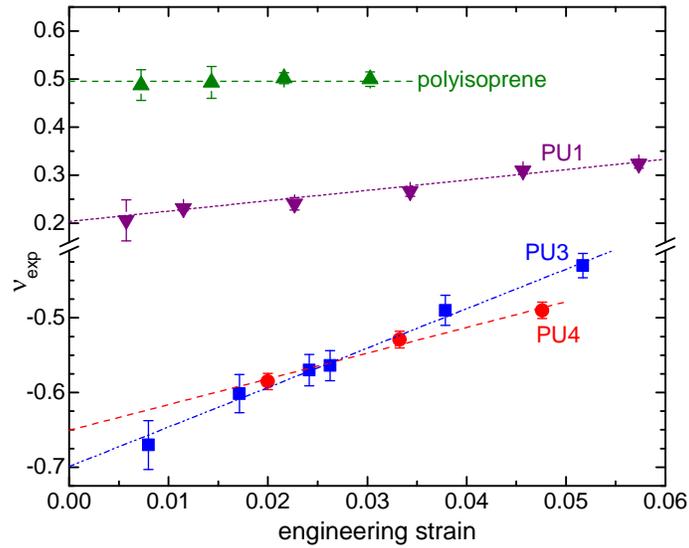

Figure 1. Directly measured Poisson's ratio for the four materials. The data for the foams has some dependence on strain, with the value obtained by linear extrapolation to $\varepsilon = 0$ listed in Table 1.

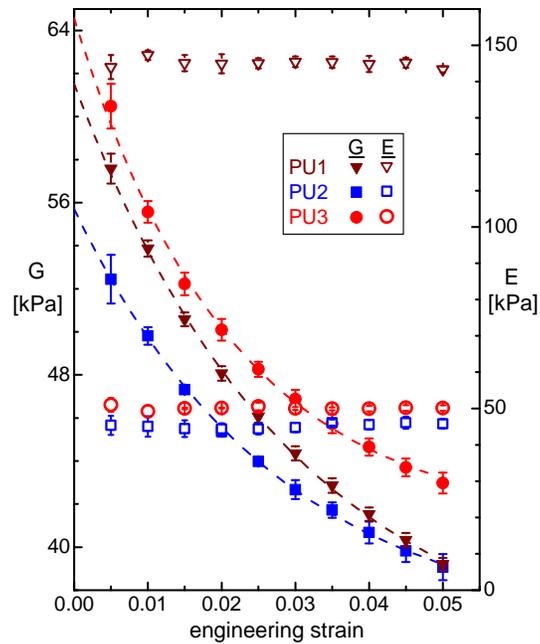

Figure 2. Engineering modulus for the three foams as a function of strain. Young's moduli are constant; fitting the shear data to a first order polynominal gives the extrapolation to zero strain listed in Table 1.